\documentclass[10pt,letterpaper]{article}
\usepackage{paperstyle}

\newcommand{\AuthorLine}{%
  Owen Tucker, BS\textsuperscript{1},
  Lily Wang, BS\textsuperscript{1},
  Harutoshi Okumura, BS\textsuperscript{1},\\
  Ruixuan Liu, PhD\textsuperscript{1},
  Li Xiong, PhD\textsuperscript{1}}
\newcommand{\AffilLine}{%
  \textsuperscript{1}Emory University, Atlanta, GA, USA}

\begin{document}

\papertitle
  {Auditing the Privacy of Synthetic Gene Expression Data: A Unified
   Weighted-Distance Framework for No-Box Membership Inference}
  {\AuthorLine}
  {\AffilLine}

\begin{paperabstract}
Synthetic gene expression data is increasingly proposed as a privacy-preserving
substitute for controlled-access genomic repositories, but its safety depends on
empirical auditing. Membership inference attacks (MIAs) provide that audit by
testing whether a patient's gene expression profile was used to train a
generative model.
We report a red-team study on synthetic gene expression data derived from bulk
RNA-seq profiles in The Cancer Genome Atlas, released by the ELSA Health 2026
Challenge. We unify five no-box
attacks under a single weighted-distance framework in which each variant differs
only in how it weights genes: uniformly (baseline), by variance, by
synthetic-versus-reference KL divergence, by spectral residualization removing
dominant principal components, and by curated pathway membership. Against a
conditional variational autoencoder, spectral residualization raises AUC from
0.8251 to 0.8951 and TPR at 1\% FPR from 0.3842 to 0.5970 on pan-cancer TCGA.
Biological pathway priors did not transfer across targets.
\end{paperabstract}

\section{Introduction}

Transcriptomic resources such as The Cancer Genome Atlas (TCGA) are among the
most reused assets in translational bioinformatics, yet the individual-level
expression profiles that make them valuable also make them identifying, and
access is therefore gated behind data use agreements and controlled-access
repositories. Synthetic data has emerged as an attractive way to relax that
gate: if a generative model can produce expression profiles that preserve
population-level biological structure without reproducing any individual
patient, the synthetic release could circulate freely where the real cohort
cannot. This promise is only as good as the empirical evidence behind it.
Generative models routinely overfit portions of their training
distribution,\cite{ward2025privacy,yeom2018privacy} recent audits show that even
differentially private generators leak when implementation deviates from
theory,\cite{annamalai2024theory,carlini2023extracting} and systematic
evaluation has found that synthetic release often fails to deliver the
anonymization its users assume.\cite{stadler2022groundhog}

Re-identification from genomic data is a long-standing concern in this field.
Homer et al.\ showed that an individual's presence in a study cohort could be
resolved from pooled allele frequencies alone,\cite{homer2008resolving} a result
that reshaped data-sharing policy at NIH, and the subsequent literature has
mapped a wide range of routes by which genetic data can be
breached.\cite{erlich2014routes} Expression data specifically is not exempt:
individual SNP genotypes can be predicted from gene expression
profiles,\cite{schadt2012bayesian} and phenotype--genotype linking attacks
quantifiably leak individual information from functional genomics
data.\cite{harmanci2016quantification} Bulk RNA-seq is therefore identifying in
its own right --- which is precisely the reason synthetic expression release is
attractive, and the reason it needs auditing rather than assurance.

Membership inference attacks (MIAs) are the standard instrument for that
evidence. An MIA asks whether a specific record was part of the data used to
train a model,\cite{shokri2017membership} and in the synthetic-data setting a
successful attack means a released cohort can be linked back to the patients who
contributed to it. MIAs are conventionally organized by the adversary's access
level: \emph{white-box} attacks exploit internal signals such as gradients or
likelihoods, \emph{black-box} attacks query the model, and \emph{no-box} attacks
see only the released synthetic data and, optionally, a public reference cohort.
The no-box setting is both the weakest for the attacker and the one that matches
how synthetic genomic data is actually shared, which makes the attainable
no-box attack strength the operationally meaningful privacy bound.

The ELSA Health 2026 Challenge instantiates exactly this audit on bulk RNA-seq
from TCGA. A blue team trains four generators --- a per-subtype multivariate
normal (MVN), a conditional variational autoencoder (CVAE), a noisy diffusion
model, and a differentially private probabilistic graphical model (DP-PGM) ---
on real patient data and releases synthetic samples; a red team must then infer
which patients were in each training set. Submissions are ranked by AUC-ROC and
by true positive rate at 1\% false positive rate (TPR@1\%), the latter capturing
whether an adversary can confidently re-identify even a small number of
patients.

Established no-box baselines such as LOGAN\cite{hayes2019logan} score a candidate
by its distance to the nearest synthetic sample, weighting all genes equally.
Our key observation is that this uniform weighting is mismatched to how
generative models fail on expression data. Membership signal concentrates in
\emph{residual} features --- the features a generator cannot smooth into a
population average. Dominant principal directions of an expression matrix encode
cancer type and broad transcriptional programs and are reproduced faithfully by
any competent generator; what survives after those are removed, along with
high-variance genes and genes whose synthetic distribution diverges from a
reference, is where per-patient memorization has room to appear.

This paper makes four contributions. First, we unify a family of no-box attacks
into a single \emph{weighted-distance} framework in which every variant is the
same cosine-distance score and differs only in a non-negative gene weight
vector. Second, we contribute a tuned no-box baseline --- cosine distance with
25th-percentile normalization, obtained by systematic search over distance
functions and normalizers --- which serves both as the trivial uniform-weight
instantiation and as a non-trivial reference for every other variant. Third, we
instantiate four notions of residual feature within the framework (spectral,
statistical, surrogate, and biological) and compose them with subtype-stratified
candidate pools. Fourth, we report the resulting empirical ranking --- among
standalone instantiations, spectral $>$ statistical $>$ surrogate $>$
biological --- together with the negative result that curated pathway priors did
not transfer across attack targets, a finding of direct relevance to
bioinformatics practitioners who might expect biological knowledge to sharpen a
privacy audit.

\section{Methods}

\subsection{Datasets and target generators}

We evaluate on the two cohorts released by the ELSA Health Challenge, both
derived from The Cancer Genome Atlas,\cite{weinstein2013tcga} and both
preprocessed by the organizers with the DESeq2 variance-stabilizing
transformation\cite{love2014deseq2} over the 978 L1000 landmark
genes.\cite{subramanian2017l1000}

\begin{itemize}
  \item \textbf{TCGA-BRCA} comprises 1,089 breast cancer patients across five
        PAM50-style subtypes (Basal, Her2, LumA, LumB, Normal-like). No separate
        reference cohort is provided.
  \item \textbf{TCGA-COMBINED} comprises 4,323 patients across ten cancer types
        (breast, colorectal, esophagus, kidney, liver, lung, ovarian,
        pancreatic, prostate, skin), together with a disjoint reference cohort
        of 824 patients.
\end{itemize}

For each cohort the challenge provides four synthetic datasets, each slightly
smaller than the original (approximately 871--872 samples for TCGA-BRCA and
3,458--3,459 for TCGA-COMBINED). MVN fits a per-subtype multivariate Gaussian
and samples from it with additive noise, making it the generator whose
failure modes are easiest to characterize.\cite{pmbio_models2025} CVAE is a conditional variational
autoencoder\cite{sohn2015learning} conditioned on subtype labels via one-hot
encoding and trained to minimize reconstruction plus KL divergence
loss.\cite{pmbio_models2025} Noisy Diffusion is a class-conditional diffusion
model that generates by iteratively denoising, with privacy obtained by adding
calibrated noise to gradients during training.\cite{noisydiffusion2025} DP-PGM
learns per-gene marginals under $\varepsilon = 10$ differential
privacy\cite{dwork2006calibrating} and samples each gene
independently.\cite{chen2024biologically}

We report results against the CVAE target. This is the most informative choice
rather than an arbitrary one: as Figure~\ref{fig:pca} shows, CVAE synthetic
samples overlap the real distribution closely, whereas the remaining three
generators produce visibly more dispersed or mismatched clouds. A generator
whose output is already distinguishable from real data offers the adversary a
trivial shortcut; CVAE forces every attack to rely on per-patient deviation
alone, and therefore gives the tightest read on how much individual-level signal
actually survives generation. Results on MVN, Noisy Diffusion, and DP-PGM are
deferred to future work.

\subsection{Baseline no-box attacks}

The challenge starter kit\cite{pmbio2025} provides six baseline MIA methods.
Four are density- or distance-based: Monte Carlo counts synthetic samples within
a fixed radius of the candidate;\cite{hilprecht2019montecarlo} GAN-Leaks
scores by minimum distance to any synthetic sample, and GAN-Leaks Calibrated
normalizes that distance against a reference set;\cite{chen2020ganleaks} and
LOGAN-D1 takes a normalized difference of reference and synthetic
distances.\cite{hayes2019logan} DOMIAS-KDE estimates a density ratio in
PCA-compressed space.\cite{vanbreugel2023domias} Two further baselines train a
logistic regression or random forest classifier on the synthetic data and use
the maximum predicted class probability as a membership score.

After evaluating and tuning hyperparameters across all six, we found that
replacing the Euclidean distance in LOGAN with cosine distance dominated the
rest, and we use that variant as the strongest baseline throughout.

\subsection{Preliminary observations}
\label{sec:prelim}

Three empirical properties of the released CVAE data motivate our attack design.

\begin{figure}[!tb]
  \centering
  \includegraphics[width=0.70\textwidth]{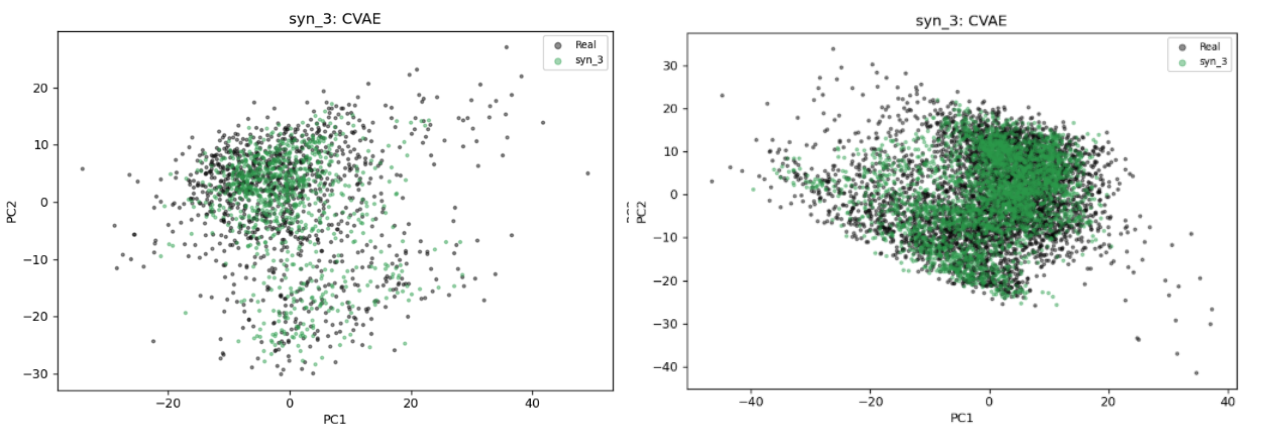}
  \caption{PCA projection comparing real (black) and CVAE synthetic data for
  TCGA-BRCA (left) and TCGA-COMBINED (right). The synthetic cloud overlaps the
  real distribution closely, making membership inference against this generator
  substantially harder than against the remaining three targets.}
  \label{fig:pca}
\end{figure}

\textbf{(i) Synthetic samples lie close to the real distribution.} In PCA space,
CVAE synthetic samples overlap the real distribution on both cohorts
(Figure~\ref{fig:pca}), so the attack difficulty we report reflects the task
rather than a weak generator.

\begin{figure}[!tb]
  \centering
  \includegraphics[width=0.70\textwidth]{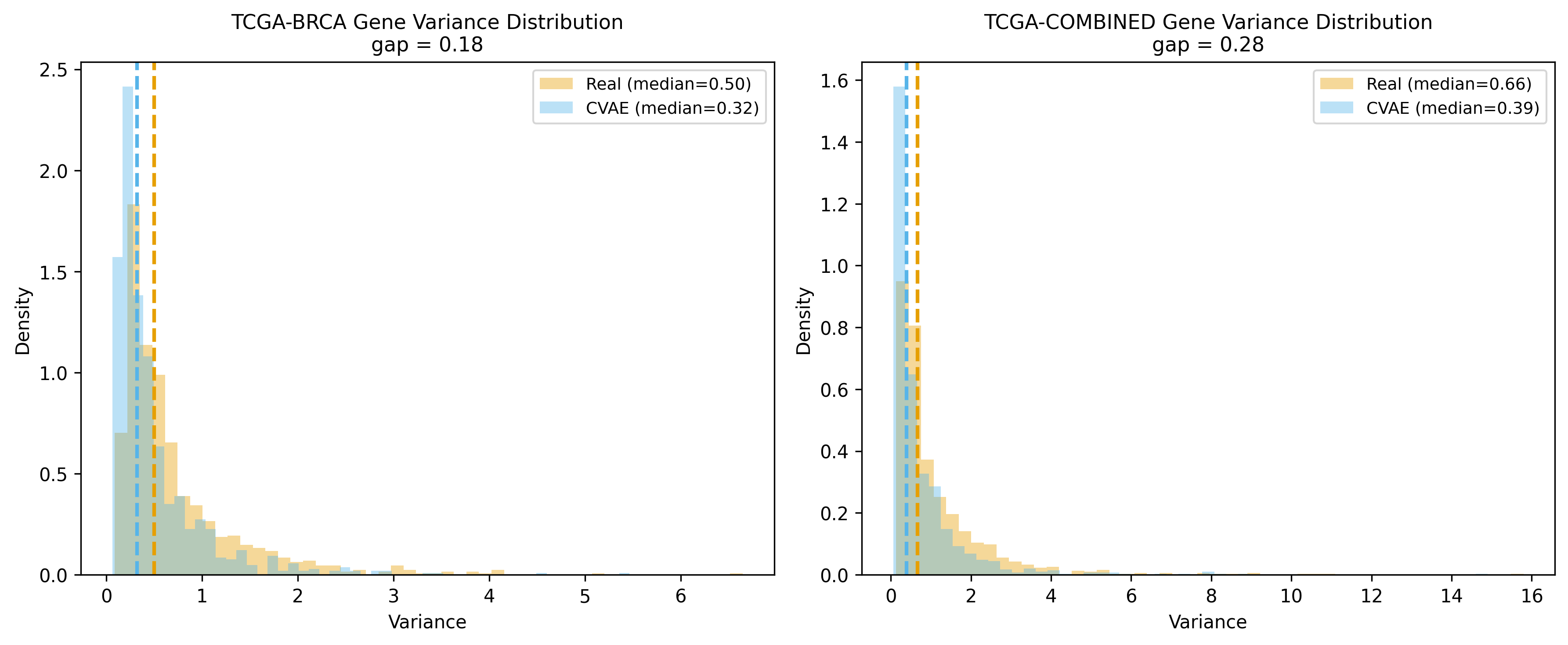}
  \caption{Per-gene variance distribution for real and CVAE synthetic data,
  TCGA-BRCA (left) and TCGA-COMBINED (right). The synthetic distribution is
  shifted toward lower variance on both cohorts, and the median gap widens with
  cohort size and heterogeneity.}
  \label{fig:variance}
\end{figure}

\textbf{(ii) Memorization signal lives in residual features.} Per-gene variance
is systematically compressed in the CVAE output --- a gap of 0.18 on TCGA-BRCA
and 0.28 on TCGA-COMBINED (Figure~\ref{fig:variance}) --- consistent with the
model smoothing its latent space toward population averages. At the same time,
the dominant variation in both real and synthetic data encodes population-level
structure: on TCGA-COMBINED the first 50 principal components explain 85\% of
variance and the first 100 explain 91\%. Together these argue that membership
signal concentrates in features that resist being smoothed into a population
average, and therefore force the generator to memorize.

\begin{figure}[!tb]
  \centering
  \begin{minipage}[t]{0.47\textwidth}
    \centering
    \includegraphics[width=\textwidth]{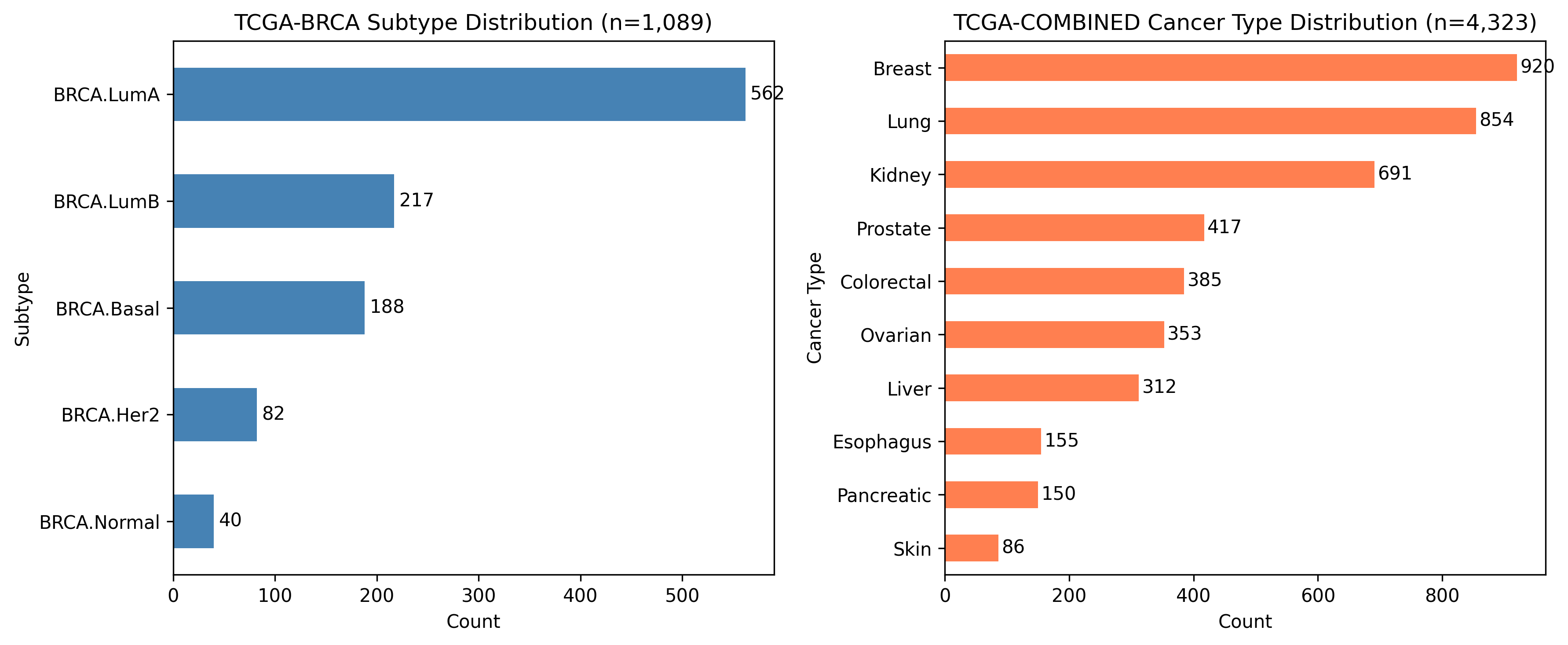}
    \caption{Subtype distribution for TCGA-BRCA (left) and cancer type
    distribution for TCGA-COMBINED (right).}
    \label{fig:subtype}
  \end{minipage}\hfill
  \begin{minipage}[t]{0.50\textwidth}
    \centering
    \includegraphics[width=\textwidth]{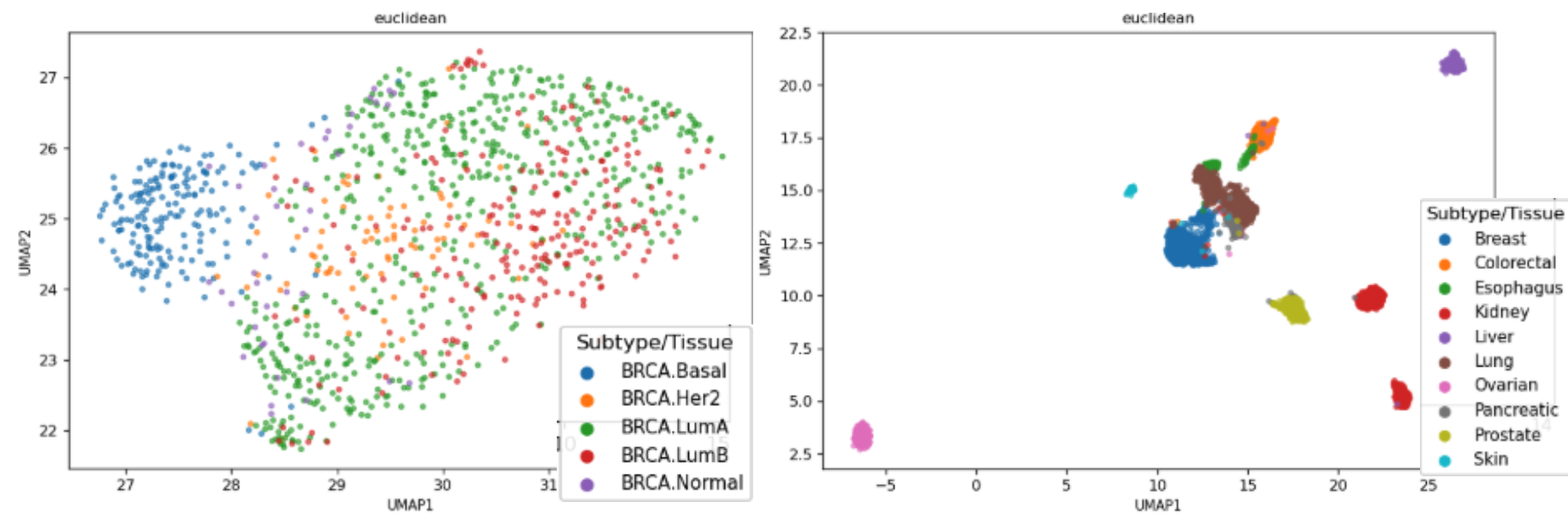}
    \caption{UMAP\cite{mcinnes2018umap} of real data by subtype (TCGA-BRCA
    left, TCGA-COMBINED right).}
    \label{fig:umap}
  \end{minipage}
\end{figure}

\textbf{(iii) Both cohorts exhibit substantial class imbalance.} In TCGA-BRCA,
LumA dominates with 562 samples (52\%) while Normal-like has only 40 (4\%)
(Figure~\ref{fig:subtype}); minority subtypes with fewer training samples are
more susceptible to generator overfitting. As Figure~\ref{fig:umap} shows,
TCGA-BRCA subtypes overlap heavily in expression space --- all five are breast
cancer variants --- whereas TCGA-COMBINED cancer types form tight, well-separated
clusters. This motivates restricting membership scores to same-subtype synthetic
samples; our expectation going in was that the restriction would pay off where
classes are biologically distinct, which the results overturn.

\subsection{A unified weighted-distance framework}
\label{sec:framework}

\textbf{Distance-based baseline.} For each candidate $x \in D_\text{mia}$ we
compute the distance to its nearest synthetic neighbor and to its nearest
reference neighbor,
$d_\text{syn}(x) = \min_{y \in D_\text{syn}} d(x, y)$ and
$d_\text{ref}(x) = \min_{y \in D_\text{ref}} d(x, y)$.
Each distance is rescaled by a percentile of its population across all
candidates, and the membership score is the difference:
\begin{equation}
  s_\text{baseline}(x)
  = \frac{d_\text{ref}(x)}{P_c(d_\text{ref})}
  - \frac{d_\text{syn}(x)}{P_c(d_\text{syn})}.
  \label{eq:logan-baseline}
\end{equation}
A higher score means the candidate sits closer to $D_\text{syn}$ than to
$D_\text{ref}$, indicating it is more likely to have been in the generator's
training set.

\textbf{Tuned baseline configuration.} Searching over distance functions and
normalizers, we found cosine distance
$d_\text{cos}(u, v) = 1 - \langle u, v \rangle / (\|u\|\,\|v\|)$
with the 25th percentile ($c = 25$) as the scaling factor to be the strongest
configuration. We denote this tuned instantiation $s_\text{best}$ and use it as
the reference point for every weighting strategy below; treating a tuned rather
than a default baseline as the comparison point avoids overstating the gains
attributable to weighting.

\textbf{From baseline to framework.} The preliminary observations above argue
that membership signal is not uniform across genes. We therefore generalize
Eq.~(\ref{eq:logan-baseline}) by replacing the unweighted cosine with a weighted
version parameterized by a non-negative weight vector
$w \in \mathbb{R}_{\geq 0}^{p}$, $p = 978$:
\begin{equation}
  d_w(u, v)
  = 1 - \frac{\sum_i w_i\, u_i\, v_i}
             {\sqrt{\sum_i w_i\, u_i^2}\,\sqrt{\sum_i w_i\, v_i^2}},
  \label{eq:weighted-cosine}
\end{equation}
leaving the percentile normalization unchanged. The tuned baseline is the
trivial case $w = \mathbf{1}$. Every strategy in the next section is a different
choice of $w$, so the framework isolates the one design decision that
distinguishes them: where to place weight on the features the generator fails to
smooth.

\subsection{Residual-based weighting strategies}
\label{sec:residual}

We present five instantiations. Four set $w$ in feature space --- three in the
gene basis (variance, KL divergence, pathway membership) and one in the PCA
basis (spectral residualization); the fifth applies the same residual-feature
logic to a surrogate-model signal.

\subsubsection{PCA residualization (spectral residual)}

The top principal components of expression data encode population-level
structure --- cancer type and major transcriptional programs --- rather than
per-patient memorization.\cite{way2020biobombe} Let $v_1, \ldots, v_p$ be the
eigenvectors of the covariance of $[D_\text{syn};\, D_\text{ref}]$ sorted by
decreasing variance explained. We \emph{residualize} each sample by zeroing its
projection onto the top $k$ components,
\begin{equation}
  \tilde{x} = x - \sum_{j=1}^{k} (x^{\top} v_j)\, v_j,
  \label{eq:residualize}
\end{equation}
and apply Eq.~(\ref{eq:logan-baseline}) to the residualized triple
$(\tilde{D}_\text{mia}, \tilde{D}_\text{syn}, \tilde{D}_\text{ref})$. In
framework terms this sets $w_j = \mathbf{1}[j > k]$ in the PCA basis. We use
$k = 100$ on TCGA-COMBINED with the 75th percentile in place of $P_{25}$, which
we found better calibrated for the rescaled residual distances. Critically, this
is residualization, not projection: all $p$ dimensions are retained. Hard
projection into a $d$-dimensional PCA subspace consistently degraded performance
in preliminary experiments (AUC $= 0.78$ at $d = 64$) because projection
permanently discards information. The scaler is fit on $D_\text{syn}$ alone,
anchoring the feature scale to the synthetic distribution, and the basis is
computed jointly without using any membership labels.

\subsubsection{Variance weighting (statistical residual)}

A high-variance gene requires the generator to learn a complex distribution and
is more prone to overfitting, whereas a low-variance gene is easy to reproduce
by tracking the population average. We weight each gene by its variance in the
synthetic data,
\begin{equation}
  w_i = \sigma^2_{\text{syn},i},
  \label{eq:var-weights}
\end{equation}
normalized so the average weight is 1.

\subsubsection{KL-divergence weighting (statistical residual)}

Variance ignores mean shifts and overall distribution mismatch. We refine it by
weighting each gene with the KL divergence between its synthetic and reference
per-gene distributions, approximated as Gaussians:
\begin{equation}
\begin{split}
  w_i &= \mathrm{KL}\!\left(\mathcal{N}(\mu_{\text{syn},i}, \sigma^2_{\text{syn},i})
        \,\big\|\, \mathcal{N}(\mu_{\text{ref},i}, \sigma^2_{\text{ref},i})\right)\\
      &= \log\frac{\sigma_{\text{ref},i}}{\sigma_{\text{syn},i}}
        + \frac{\sigma^2_{\text{syn},i} + (\mu_{\text{syn},i} - \mu_{\text{ref},i})^2}
               {2\sigma^2_{\text{ref},i}} - \tfrac{1}{2},
\end{split}
\label{eq:kl-weights}
\end{equation}
clipping negative values arising from numerical artifacts to zero and
normalizing to mean one. TCGA-BRCA provides no clean reference cohort, so we
substitute $D_\text{mia}$ for $D_\text{ref}$; this slightly contaminates the
weights, since $D_\text{mia}$ mixes members and non-members.

\subsubsection{MinK-style surrogate scoring (surrogate residual)}

The strategies above weight the LOGAN distance signal. We additionally apply the
residual-feature logic to a different signal: the per-gene reconstruction error
of a surrogate generator --- a CVAE replicating the blue-team architecture,
trained on $D_\text{syn}$ as a stand-in for the inaccessible target. For each
candidate $x$, let $r_i(x) = (x_i - \hat{x}_i)^2$ be the surrogate's per-gene
squared error. Following MinK for language-model membership
inference,\cite{shi2023detecting} we score each candidate by the negated mean
reconstruction error over its hardest genes,
\begin{equation}
  s_\text{MinK}(x) = -\frac{1}{|S(x)|}\sum_{i \in S(x)} r_i(x),
  \quad
  S(x) = \{\, i : r_i(x) \text{ among the top-}10\% \text{ largest} \,\},
  \label{eq:mink}
\end{equation}
which is a binary, candidate-dependent weighting on the reconstruction-error
vector. The intuition is that the surrogate is trained on $D_\text{syn}$, which
was itself generated from training-set patients, so the target's memorization
leaks through: for members, even the hardest genes reconstruct reasonably well,
whereas for non-members they remain difficult. We also fuse with the tuned
baseline by z-score normalization,
$s_\text{MinK+best}(x) = z(s_\text{MinK}(x)) + z(s_\text{best}(x))$.

\subsubsection{Biologically informed gene selection (biological residual)}

The four strategies above are agnostic to gene identity. We also test a
biological notion of residual: a binary mask restricting attention to a curated
pathway gene set $\mathcal{P}$. Using g:Profiler g:GOSt
enrichment\cite{raudvere2019gprofiler} over the 978 landmark genes against
KEGG\cite{kanehisa2000kegg} and Reactome,\cite{jassal2020reactome} we performed
an exhaustive combinatorial search
of pathway combinations, evaluating five attack variants per combination across
86,098 configurations. The best-performing combination --- 63 genes spanning
autophagy, ESR-mediated signaling, and nuclear receptor signaling --- defines
$\mathcal{P}$. We report two configurations. \emph{Pathway unweighted} restricts
the cosine distance to $\mathcal{P}$ via
\begin{equation}
  w_i = \mathbf{1}[i \in \mathcal{P}],
  \label{eq:pathway-weights}
\end{equation}
zeroing the remaining 915 genes. \emph{Full KL + pathway} z-score fuses this
pathway-restricted score with the KL-weighted full-gene score,
$s = z(s_\text{KL}) + z(s_\text{pathway})$, on the hypothesis that statistical
divergence and biological relevance carry complementary signal.

\subsection{Subtype-stratified candidate pools}

Orthogonal to the choice of weights, the challenge's release of synthetic
subtype labels enables restricting the candidate pool itself rather than
reweighting features. For a candidate of subtype $c$, \emph{subtype-stratified}
replaces $D_\text{syn}$ in Eq.~(\ref{eq:logan-baseline}) with the same-subtype
subset $D_\text{syn}^{(c)}$ while leaving $D_\text{ref}$ unchanged;
\emph{fully stratified} additionally restricts $D_\text{ref}$ to
$D_\text{ref}^{(c)}$, isolating membership signal within each cancer type's
expression subspace. Both are candidate-pool restrictions rather than feature
weightings, and we report them as a separate axis.

\subsection{Evaluation}

We report the two challenge metrics: AUC-ROC and TPR at FPR $= 0.01$.
Distance-based attacks are evaluated under a LiRA-style shadow
framework\cite{carlini2022firstprinciples,zarifzadeh2024lowcost} with $M$ shadow CVAE generators trained on
random 50\% subsets of the data (TCGA-BRCA averaged over 10 targets;
TCGA-COMBINED over 3), and membership is inferred via a likelihood-ratio test
comparing in- versus out-score distributions. Methods marked $\dagger$ in
Table~\ref{tab:results} are reported under a local-oracle split (single seed;
50/30/20 member / non-member / reference) for which the shadow framework was not
run, and should be read as indicative rather than as directly comparable to the
shadow-framework rows. Implementation code is available at
\url{https://github.com/OwenTucker/ELSA_REDSIGMA/}.

\section{Results}

Table~\ref{tab:results} reports no-box MIA performance for all instantiations of
the framework on TCGA-BRCA and TCGA-COMBINED against the CVAE generator,
alongside the unweighted baselines.

\begin{table}[!t]
\centering
\caption{Membership inference results on TCGA-BRCA and TCGA-COMBINED against the
CVAE generator. Shadow-framework results are averaged over 10 targets (BRCA) and
3 targets (COMBINED); $^\dagger$ denotes a local-oracle simulation at a single
seed. Best result per column in bold.}
\label{tab:results}
\vspace{2pt}
{\fontsize{9}{11}\selectfont
\begin{tabular}{llcccc}
\toprule
& & \multicolumn{2}{c}{\textbf{TCGA-BRCA}} & \multicolumn{2}{c}{\textbf{TCGA-COMBINED}}\\
\cmidrule(lr){3-4}\cmidrule(lr){5-6}
\textbf{Category} & \textbf{Method} & \textbf{AUC-ROC} & \textbf{TPR@1\%}
                  & \textbf{AUC-ROC} & \textbf{TPR@1\%}\\
\midrule
\multirow{3}{*}{Baseline}
  & LOGAN cosine $P_{25}$, $k=1$      & 0.7166 & 0.2003 & 0.8251 & 0.3842\\
  & GAN-Leaks cosine $k=1$            & 0.7166 & 0.2003 & 0.8200 & 0.3542\\
  & DOMIAS-KDE                        & 0.7166 & 0.2003 & 0.4828 & 0.0130\\
\midrule
\multirow{2}{*}{Feature weighting}
  & Weighted cosine (variance)        & 0.7214 & 0.1918 & 0.7775 & 0.3092\\
  & Weighted cosine (KL)              & \textbf{0.7231} & 0.1853 & 0.8295 & 0.3691\\
\midrule
\multirow{2}{*}{PCA residualization}
  & Minus top-50 PCs$^\dagger$        & 0.7140 & 0.1900 & 0.8769 & 0.5294\\
  & Minus top-100 PCs$^\dagger$       & 0.7170 & 0.1490 & \textbf{0.8951} & \textbf{0.5970}\\
\midrule
\multirow{4}{*}{Pathway + subtype}
  & Pathway unweighted                & 0.6794 & 0.1689 & 0.6972 & 0.1123\\
  & Full KL + pathway                 & 0.7025 & 0.1872 & 0.7249 & 0.1169\\
  & Subtype-stratified$^\dagger$      & 0.7049 & 0.3346 & 0.8269 & 0.3766\\
  & Fully stratified$^\dagger$        & 0.7092 & \textbf{0.3456} & 0.8281 & 0.3752\\
\midrule
\multirow{2}{*}{MinK surrogate}
  & MinK surrogate (10\%)             & 0.6505 & 0.2089 & 0.8063 & 0.4171\\
  & MinK surrogate + LOGAN            & 0.6968 & 0.2232 & 0.8403 & 0.4815\\
\bottomrule
\end{tabular}}
\end{table}

\textbf{Spectral residuals provide the strongest signal, but only on
TCGA-COMBINED.} Removing the top 100 principal components lifts COMBINED AUC
from the baseline 0.8251 to 0.8951 and TPR@1\% from 0.3842 to 0.5970 --- the
largest single-method improvement we observed. The Spearman correlation between
raw and residualized scores is moderate ($\rho = 0.77$), confirming that
residualization re-ranks candidates rather than merely rescaling them. On
TCGA-BRCA the same procedure does nothing: AUC moves from 0.7166 to 0.7170 and
TPR@1\% falls from 0.2003 to 0.1490. We attribute the split to the reference
cohort. The PCA basis is estimated from $[D_\text{syn}; D_\text{ref}]$, and
TCGA-BRCA has no clean $D_\text{ref}$, so the basis is contaminated by the very
candidates being scored and the components removed are not purely
population-level.

\textbf{KL weighting is the most consistent improvement in AUC.} KL weighting
beats the tuned baseline on both cohorts ($0.7166 \to 0.7231$ on BRCA;
$0.8251 \to 0.8295$ on COMBINED). Variance weighting helps marginally on BRCA
but degrades COMBINED substantially ($0.8251 \to 0.7775$). Neither improves
TPR@1\%: both weightings trade a small AUC gain for a loss at the strict
operating point (KL: $0.2003 \to 0.1853$ on BRCA, $0.3842 \to 0.3691$ on
COMBINED), so per-gene reweighting sharpens average-case discrimination without
helping the adversary who needs high-confidence hits.

\textbf{Surrogate residuals are complementary, not competitive.} Standalone MinK
is weaker than the baseline on both cohorts, but its z-score fusion with LOGAN
raises COMBINED AUC to 0.8403 --- the gain comes from fusion, not from MinK
alone.

\textbf{Subtype stratification sharpens the low-FPR regime --- on the wrong
cohort.} Both stratified variants leave AUC near baseline but raise TPR@1\% on
TCGA-BRCA from 0.2003 to 0.3346 and 0.3456, the best low-FPR results on that
cohort and a $72\%$ relative gain. This runs against the expectation set out in
the preliminary observations: we predicted stratification would pay off on
TCGA-COMBINED, where cancer types are well separated in expression space, yet
there it is flat to slightly negative ($0.3842 \to 0.3766$ and $0.3752$). The
reversal is informative. Where classes are already well separated, the full
synthetic pool contains few cross-type near neighbors, so restricting it removes
little; where classes overlap heavily, as the five breast subtypes do, the
nearest synthetic neighbor of a candidate is often from another subtype, and
excluding those neighbors is what exposes the within-subtype memorization
signal.

\textbf{Biological residuals do not transfer.} Despite the exhaustive search over
86,098 pathway configurations, the best pathway-restricted attack underperforms
the full-gene baseline on both cohorts (0.6794 and 0.6972 AUC), and fusing it
with the KL-weighted score does not improve on KL alone. The winning pathway
combination at the discovery seed did not generalize to other targets.

\section{Discussion}

\textbf{The framework converts a list of tricks into a testable claim.} Casting
the tuned baseline, PCA residualization, variance and KL weighting, MinK, and
pathway selection as instantiations of a single weighted-distance score isolates
the design decision that matters: where weight is placed on the features the
generator fails to smooth. Among standalone instantiations the resulting
ranking --- spectral $>$ statistical $>$ surrogate $>$ biological --- is a
falsifiable statement about which notion of residual is most concentrated in a
CVAE trained on TCGA, not merely an aggregate leaderboard position. The ranking
is not invariant to fusion: z-scoring MinK with the tuned baseline lifts the
surrogate above the statistical strategies on TCGA-COMBINED ($0.8403$ versus
$0.8295$), which is what one expects if the two signals are complementary rather
than substitutable.

\textbf{Why spectral residuals win.} Removing dominant principal components
strips exactly the structure a conditional generator is explicitly trained to
reproduce --- cancer type and major transcriptional programs --- leaving the
per-patient deviation that the model had no incentive to smooth. That
residualization outperforms hard projection into a low-dimensional subspace
(0.8951 versus 0.78 at $d = 64$) reinforces the point: the goal is to
\emph{deweight} population structure, not to discard the dimensions in which
individual signal is expressed.

\textbf{Why biological priors fail, and why that matters here.} The negative
pathway result is, for a translational bioinformatics audience, the more
actionable finding. Pathway membership encodes what is biologically meaningful
\emph{across} patients, which is close to the opposite of what a membership
attack needs: features that are idiosyncratic \emph{to} a patient. Curated
pathways therefore select genes whose behavior the generator is best positioned
to learn. The practical implication is that privacy audits of synthetic omics
data should not be assumed adequate because they incorporate domain knowledge;
in our experiments the released data itself was a more reliable guide to where
leakage lives than any external biological prior.

\textbf{Reference data is the limiting resource.} TCGA-BRCA's smaller scale and
absent clean reference cohort degrade every comparison-based strategy, while
TCGA-COMBINED's 824-patient reference enables better KL weight estimates, better
baseline normalization, and --- decisively --- a clean PCA basis. The reference
cohort is what separates our strongest result from our weakest: the identical
residualization procedure gains $0.07$ AUC on COMBINED and nothing on BRCA. This
accounts for most of the cross-cohort gap and has a direct operational reading:
the privacy risk of a synthetic release is not a fixed property of the generator
but depends on what auxiliary population data an adversary can obtain. As public
expression compendia grow, the same release becomes progressively easier to
attack, and a release audited as safe today is not thereby safe indefinitely.

\textbf{Implications for synthetic data release.} A TPR@1\% of 0.597 means that
at a threshold flagging only $1\%$ of non-members, an adversary still recovers
roughly $60\%$ of the patients who were actually in the training cohort. For a
release derived from cancer patients, that is not a residual risk. The CVAE examined here is a
non-private generator, so this is a measurement of an unprotected baseline
rather than a break of a privacy guarantee --- but it quantifies how far
``looks like the population'' is from ``protects the individual,'' and argues that
distributional fidelity checks are not a substitute for adversarial auditing
before a synthetic omics dataset is released.

\textbf{Limitations.} Our results characterize a single generator family on a
single data modality. Rows marked $\dagger$ in Table~\ref{tab:results} come from
a single-seed local-oracle split rather than the shadow framework, so the PCA
residualization and stratification gains, while large, rest on a weaker
evaluation protocol than the shadow-framework rows and warrant confirmation.
KL weights on TCGA-BRCA are estimated from $D_\text{mia}$, which mixes members
and non-members and mildly contaminates that column. Finally, the pathway search
was conducted at a single discovery seed; a target-aware selection procedure
might yet recover signal we could not.

\textbf{Future work.} Natural next steps are rescuing PCA residualization on
TCGA-BRCA via pseudo-reference construction, surrogate generators trained jointly on
$D_\text{syn}$ and $D_\text{ref}$, target-aware rather than seed-fixed pathway
selection, and extension of the framework to the remaining three challenge
generators --- particularly the two differentially private ones, where the
framework can be used to measure the empirical gap between a claimed
$\varepsilon$ and attainable no-box attack strength.

\section{Conclusion}

We unified five no-box membership inference strategies for synthetic gene
expression under a single weighted-distance framework, organized by the kind of
residual feature each strategy emphasizes: spectral, statistical, surrogate, and
biological. Alongside the framework we contribute a tuned no-box baseline that
serves both as the trivial uniform-weight instantiation and as a non-trivial
reference for comparison. Spectral residuals do most of the work where a clean
reference cohort exists, lifting AUC from 0.8251 to 0.8951 and TPR@1\% from
0.3842 to 0.5970 on pan-cancer TCGA by stripping the population-level structure
that masks per-patient memorization, and doing nothing on TCGA-BRCA, where no
such cohort is available. KL divergence is the most reliable per-gene weighting
in AUC, though no weighting helped at the strict operating point. Two results
ran against our expectations: curated biological pathways did not transfer
across targets, and subtype stratification paid off on the cohort whose classes
overlap rather than the one where they separate cleanly. For
translational bioinformatics, the practical conclusion is that leakage in
synthetic omics releases is best located in the released data itself rather than
through external biological priors, and that the strength of such an audit is
bounded chiefly by the reference cohort available to the auditor.

\section{Generative AI Disclosure}

A large language model (Claude, Anthropic) was used in manuscript preparation:
formatting, and polishing the Introduction and
Discussion. The four audited generators were trained and released by the ELSA
Health 2026 Challenge organizers; we trained one surrogate conditional
variational autoencoder on their synthetic output, as described in Methods. All
experiments, analyses, and numerical results were produced and verified by the
authors.

\begin{paperreferences}{99}

\bibitem{ward2025privacy}
Ward J, Wang CH, Cheng G. Privacy auditing synthetic data release through local
likelihood attacks. arXiv:2508.21146 [preprint]. 2025.

\bibitem{yeom2018privacy}
Yeom S, Giacomelli I, Fredrikson M, Jha S. Privacy risk in machine learning:
analyzing the connection to overfitting. In: 2018 IEEE 31st Computer Security
Foundations Symposium (CSF). IEEE; 2018. p. 268--82.

\bibitem{annamalai2024theory}
Annamalai MSMS, Ganev G, De Cristofaro E. ``What do you want from theory
alone?'' Experimenting with tight auditing of differentially private synthetic
data generation. arXiv:2405.10994 [preprint]. 2024.

\bibitem{carlini2023extracting}
Carlini N, Hayes J, Nasr M, Jagielski M, Sehwag V, Tram\`er F, et al.
Extracting training data from diffusion models. In: Proceedings of the 32nd
USENIX Security Symposium. 2023.

\bibitem{stadler2022groundhog}
Stadler T, Oprisanu B, Troncoso C. Synthetic data --- anonymisation groundhog
day. In: Proceedings of the 31st USENIX Security Symposium. 2022. p. 1451--68.

\bibitem{homer2008resolving}
Homer N, Szelinger S, Redman M, Duggan D, Tembe W, Muehling J, et al. Resolving
individuals contributing trace amounts of DNA to highly complex mixtures using
high-density SNP genotyping microarrays. PLoS Genet. 2008;4(8):e1000167.

\bibitem{erlich2014routes}
Erlich Y, Narayanan A. Routes for breaching and protecting genetic privacy. Nat
Rev Genet. 2014;15(6):409--21.

\bibitem{schadt2012bayesian}
Schadt EE, Woo S, Hao K. Bayesian method to predict individual SNP genotypes
from gene expression data. Nat Genet. 2012;44(5):603--8.

\bibitem{harmanci2016quantification}
Harmanci A, Gerstein M. Quantification of private information leakage from
phenotype-genotype data: linking attacks. Nat Methods. 2016;13(3):251--6.

\bibitem{shokri2017membership}
Shokri R, Stronati M, Song C, Shmatikov V. Membership inference attacks against
machine learning models. In: 2017 IEEE Symposium on Security and Privacy (SP).
IEEE; 2017. p. 3--18.

\bibitem{hayes2019logan}
Hayes J, Melis L, Danezis G, De Cristofaro E. LOGAN: membership inference
attacks against generative models. Proc Priv Enhancing Technol.
2019;2019(1):133--52.

\bibitem{weinstein2013tcga}
Weinstein JN, Collisson EA, Mills GB, Shaw KRM, Ozenberger BA, Ellrott K, et al.
The Cancer Genome Atlas Pan-Cancer analysis project. Nat Genet.
2013;45(10):1113--20.

\bibitem{love2014deseq2}
Love MI, Huber W, Anders S. Moderated estimation of fold change and dispersion
for RNA-seq data with DESeq2. Genome Biol. 2014;15(12):550.

\bibitem{subramanian2017l1000}
Subramanian A, Narayan R, Corsello SM, Peck DD, Natoli TE, Lu X, et al. A next
generation connectivity map: L1000 platform and the first 1,000,000 profiles.
Cell. 2017;171(6):1437--52.

\bibitem{pmbio_models2025}
PMBio. Baseline generative models for the CAMDA 2026 Health Privacy Challenge
[Internet]. GitHub; 2025 [cited 2026 Sep 3]. Available from:
\url{https://github.com/PMBio/Health-Privacy-Challenge/tree/main/src/generators/models}

\bibitem{sohn2015learning}
Sohn K, Lee H, Yan X. Learning structured output representation using deep
conditional generative models. In: Advances in Neural Information Processing
Systems. Vol. 28. 2015.

\bibitem{noisydiffusion2025}
Kreuer J, Ouaari S. Noisy diffusion: a diffusion-based generative model for gene
expression data [Internet]. ELSA Benchmarks; 2025 [cited 2026 Sep 3]. Available
from: \url{https://benchmarks.elsa-ai.eu/}

\bibitem{dwork2006calibrating}
Dwork C, McSherry F, Nissim K, Smith A. Calibrating noise to sensitivity in
private data analysis. In: Theory of Cryptography Conference (TCC). Springer;
2006. p. 265--84.

\bibitem{chen2024biologically}
Chen D, Oestreich M, Afonja T, Kerkouche R, Becker M, Fritz M. Towards
biologically plausible and private gene expression data generation. Proc Priv
Enhancing Technol. 2024.

\bibitem{pmbio2025}
PMBio. Health-Privacy-Challenge: starter kit for the CAMDA 2026 Health Privacy
Challenge [Internet]. GitHub; 2025 [cited 2026 Sep 3]. Available from:
\url{https://github.com/PMBio/Health-Privacy-Challenge}

\bibitem{hilprecht2019montecarlo}
Hilprecht B, H\"arterich M, Bernau D. Monte Carlo and reconstruction membership
inference attacks against generative models. Proc Priv Enhancing Technol.
2019;2019(4):232--49.

\bibitem{chen2020ganleaks}
Chen D, Yu N, Zhang Y, Fritz M. GAN-Leaks: a taxonomy of membership inference
attacks against generative models. In: Proceedings of the 2020 ACM SIGSAC
Conference on Computer and Communications Security. 2020. p. 343--62.

\bibitem{vanbreugel2023domias}
van Breugel B, Sun H, Qian Z, van der Schaar M. Membership inference attacks
against synthetic data through overfitting detection. In: Proceedings of the
26th International Conference on Artificial Intelligence and Statistics. 2023.
p. 3493--514.

\bibitem{mcinnes2018umap}
McInnes L, Healy J, Melville J. UMAP: uniform manifold approximation and
projection for dimension reduction. arXiv:1802.03426 [preprint]. 2018.

\bibitem{way2020biobombe}
Way GP, Zietz M, Rubinetti V, Himmelstein DS, Greene CS. Compressing gene
expression data using multiple latent space dimensionalities learns
complementary biological representations. Genome Biol. 2020;21(1):109.

\bibitem{shi2023detecting}
Shi W, Ajith A, Xia M, Huang Y, Liu D, Blevins T, et al. Detecting pretraining
data from large language models. arXiv:2310.16789 [preprint]. 2023.

\bibitem{raudvere2019gprofiler}
Raudvere U, Kolberg L, Kuzmin I, Arak T, Adler P, Peterson H, et al. g:Profiler:
a web server for functional enrichment analysis and conversions of gene lists
(2019 update). Nucleic Acids Res. 2019;47(W1):W191--8.

\bibitem{kanehisa2000kegg}
Kanehisa M, Goto S. KEGG: Kyoto Encyclopedia of Genes and Genomes. Nucleic Acids
Res. 2000;28(1):27--30.

\bibitem{jassal2020reactome}
Jassal B, Matthews L, Viteri G, Gong C, Lorente P, Fabregat A, et al. The
Reactome pathway knowledgebase. Nucleic Acids Res. 2020;48(D1):D498--503.

\bibitem{carlini2022firstprinciples}
Carlini N, Chien S, Nasr M, Song S, Terzis A, Tram\`er F. Membership inference
attacks from first principles. In: 2022 IEEE Symposium on Security and Privacy
(SP). IEEE; 2022. p. 1897--914.

\bibitem{zarifzadeh2024lowcost}
Zarifzadeh S, Liu P, Shokri R. Low-cost high-power membership inference attacks.
In: Proceedings of the 41st International Conference on Machine Learning. 2024.

\end{paperreferences}

\end{document}